\documentclass[twocolumn]{revtex4}
\usepackage{graphicx}
\usepackage{amsmath}
\usepackage{balance}
\begin{document}

\begin{flushleft}
 LPNHE/2006-08\\
\end{flushleft}

\newcommand{\sdot}{\!\cdot\!}
\newcommand{\SLASH}[1]{/\!\!\! #1}

\title{PSEUDOSCALAR-SCALAR  TRANSITION FORM FACTORS \\
IN COVARIANT LIGHT FRONT DYNAMICS}

\author{O.M.A. LEITNER$^\circ$\footnote{Talk given at the XXXIII International Conference on High Energy Physics, ICHEP06, Moscow, 
26 July-02 August.}}
\author{B. EL-BENNICH$^\dagger$}
\author{B. LOISEAU$^\ddagger$}

\affiliation{Laboratoire de Physique Nucl\'eaire et de Hautes \'Energies, IN2P3-CNRS \\
Universit\'e P. $\&$ M. Curie, 4 Pl. Jussieu, 75252 Paris, 
France\\
$^\circ$ E-mail: leitner@lpnhep.in2p3.fr\\
$^\dagger$ E-mail: bruno.elbennich@lpnhe.in2p3.fr\\
$^\ddagger$ E-mail: loiseau@lpnhep.in2p3.fr\\}

\author{J.P. DEDONDER$^\natural$}

\affiliation{Universit\'e Denis Diderot, GMPIB, case 7021 - F 75251 Paris, France\\
$^\natural$ E-mail: dedonder@paris7.jussieu.fr}

\begin{abstract}
In an explicitly covariant light-front formalism, we 
analyze transition form factors between pseudoscalar and scalar mesons. Application 
is performed in case of the $B \to f_0(980)$ transition in the full available 
transfer momentum range $q^2$.
\end{abstract}
\maketitle

\section{Covariant Light Front Formalism}
In the past few years, a generalization of the standard Light-Front Dynamics (LFD)
has been proposed: Covariant  Light-Front Dynamics\cite{Carbonell:1998rj} (CLFD). The formulation 
has already been succesfully applied to relativistic particle and nuclear physics and it is particularly 
useful for  describing hadrons, and all observables related to them, within the constituent quark model. 
In CLFD\cite{Carbonell:1998rj}, the state vector which describes the physical bound state is 
defined on the light-front plane given by the equation $\omega \sdot r = \sigma$, where $\omega$  
is an unspecified light-like  four vector ($ \omega^{2} = 0$) which  defines the position of the 
light-front plane and $r$ is a four vector position of the system.

\begin{figure}[b] 
\begin{center}
\includegraphics*[width=0.65\columnwidth]{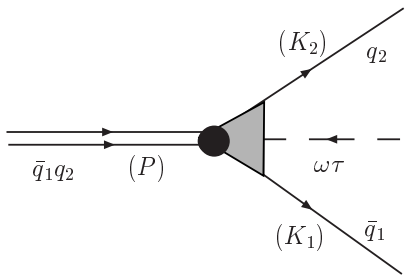}
\end{center}
\caption{Representation of the two body wave function on the light front.} 
\label{fig1} 
\end{figure}

CLFD proposes a formulation in which the evolution for a given system is expressed 
in terms of covariant expressions. 
Any four vector describing a phenomena can be transformed from one system of reference to another 
one by using a unique standard matrix which depends only on  kinematical parameters and on $\omega$. 
The particle is described by a wave function expressed in terms of Fock components of the state vector  
which respects the properties required under any transformation.  
The state vector describing a meson of momentum $p$, defined on a light-front plane characterized by 
$\omega$, is given by:
\begin{multline}\label{eq7.8}
{| p, \lambda \rangle}_{\omega} = (2 \pi)^{3/2}  \\ \int \Phi_{j_{1}\sigma_{1}j_{2}
\sigma_{2}}^{J\lambda}
(k_{1},k_{2},p,\omega \tau)a_{\sigma_{1}}^{\dagger}({\bf k}_{1})a_{\sigma_{2}}^{\dagger}({\bf k}_{2})|0
\rangle \\
 \times
\delta^{(4)}(k_{1}+k_{2}-p-\omega \tau) \exp(i \tau \sigma)2(\omega \sdot p) {\rm d}\tau  \\ \frac{{\rm d}^{3}
k_{1}}
{(2 \pi)^{3/2}\sqrt{2 \varepsilon_{k_{1}}}} \frac{{\rm d}^{3}k_{2}}
{(2 \pi)^{3/2}\sqrt{2 \varepsilon_{k_{2}}}} \ ,
\end{multline}
where $\varepsilon_{k_{i}}=\sqrt{{\bf k}_{i}^2+m_{i}^2}$ and ${\bf k}_{i}$ is the momentum of the quark $i$. 
In Eq.~(\ref{eq7.8}) $\lambda$ is the projection of the total angular momentum, $J$, of the system on the $z$ 
axis in the rest frame and $\sigma_{1}, \sigma_{2}$ are the spin projections of the 
particles 1 to 2 in the corresponding rest systems. From the delta function ensuring momentum conservation, one gets:
\begin{equation}\label{eq7.9}
\mathcal{P}= p+ \omega \tau = k_{1}+k_{2}\ .
\end{equation}
To keep track of this conservation law, a momentum, $\omega \tau$, is assigned to the spurion but there is 
no fictitious particle in the physical state vector (see Fig.~1).
We emphasize that the bound state wave function is always an off-energy shell object ($\tau \neq  0$  due 
to binding energy) and depends on the light-front orientation.  
The parameter $\tau$ is entirely determined by the on-mass shell condition for the individual constituents. 
The two-body wave function $\Phi^{J\lambda}(k_1,k_2,p,\omega \tau)$ written in Eq.~(\ref{eq7.8}) can be parametrized 
in terms of various sets of variables. We shall use in the following  the usual light-front coordinates 
$(x,\bf{R}_{\perp})$, which are defined by analogy to the equal time function in the infinite momentum frame as:
\begin{eqnarray}\label{eq7.13}
x&=& \frac{\omega \sdot k_{1} }{ \omega \sdot p}\ ,\nonumber  \\
R_{1}&=&k_{1}-xp \ ,\nonumber 
\end{eqnarray}
and where $R_1$ is decomposed in its spatial components parallel and perpendicular to the direction of the 
light-front,   $ R_{1}=(R_{0},{\bf R}_{\perp}, {\bf R}_{\|})$. 
We have by definition $R_1 \sdot 
\omega=0$, and thus $R_1^2=-{\bf R}_{\perp}^2$. 

\section{Scalar Particle in Light Front}
The explicit covariance of our approach allows us to write down the general 
structure of the two-body bound state. For a scalar particle composed of an  antiquark 
and a quark of same mass, $m$, it has the form:                       
\begin{equation}\label{eq58.1}                                                    
\Phi=\frac{1}{\sqrt{2}}\bar{u}(k_2)\left[\frac{A({\bf k}^{2})}{m}\right]v(k_1)\ ,
\end{equation}
where  $v(k_{1})$ and  ${\Bar u}(k_{2})$  are the usual antiparticle and particle Dirac spinors, and $A({\bf k}^{2})$ 
is the  scalar component of the meson wave function, $\Phi$. The expression for $A({\bf k}^{2})$ is as follows:
\begin{eqnarray}
A({\bf k}^{2}) = \frac{m}{2 \varepsilon_{k}} g({\bf k}^{2})\ ,
\end{eqnarray}
where one defines $\varepsilon_{k}=\sqrt{{\bf k}^2+m^2}.$ 
The component $g({\bf k}^{2})$ will be parametrized by a gaussian wave function written 
as $g({\bf k}^{2})=4  \pi^2 \alpha \beta  
\exp(-\beta {\bf k}^{2})$  
where $\alpha$ and $\beta$ are  parameters to be determined from experimental 
data and theoretical assumptions. In terms of the variables $(x,{\bf R}_\perp)$, we have for the relative 
momentum between two quarks of same  masses:
\begin{equation}\label{eq8.5}
{\bf k}^{2}=\frac{{\bf R}_{\perp}^{2}+m^{2}}{4x(1-x)}-m^2\ ,
\end{equation}
where the relativistic, relative momentum, ${\bf k}$, corresponds, in the frame 
where ${\bf k_1}+{\bf k_2}=\bf{0}$, to the usual relative momentum between the two particles. 

\section{Pseudoscalar-Scalar ($\boldsymbol{P \to S}$) Transition Form Factors}
In the  Covariant Light-Front Dynamics formalism, the exact transition amplitude does not depend on the light 
front orientation. However, in any approximate computation the dependence  is explicit and  we 
can parametrize this dependence  since our formalism is covariant. 
Hence, the approximate amplitude expressed in CLFD is given by the following hadronic matrix,
\begin{multline}\label{eq9.7}
\langle S(P_{2}) | \gamma^{\mu}\gamma^5| P(P_{1}) \rangle = 
(P_{1}+P_{2})^{\mu} f_{+}(q^{2})   \\ +(P_{1}-P_{2})^{\mu} f_{-}(q^{2})
+ B(q^{2})\omega^{\mu}\ ,  
\end{multline}
where $B(q^{2})$ is a non-physical form factor which has to be zero in any exact calculation. 
The last term in Eq.~(6) represents the explicit dependence of the amplitude on the light front orientation $\omega$. In order 
to extract the physical form factor $f{_{\pm}}(q^{2})$,  without any dependence on $\omega$, from the amplitude 
$\langle S(P_{2}) | J^{\mu}| P(P_{1}) \rangle$, we will proceed as  follow. 
First, we calculate the scalar products $\mathcal{X}, \mathcal{Y}$ and $ \mathcal{Z}$ which are defined by,
\begin{multline}\label{eq9.8}
\mathcal{X} = (P_1+P_2)_{\mu} \sdot \langle S(P_{2}) | J^{\mu}| P(P_{1}) \rangle =  \\
f_{+}(q^{2})  \Bigl[ 2( M_{1}^2 +  M_{2}^{2}) -q^2 \Bigr]  \\
+ f_{-}(q^{2}) (M_{1}^2-M_{2}^2) 
+ B(q^{2}) P_1 \sdot \omega  \; ( 1 +  y )\ ,
\end{multline}
\vskip -0.6cm
\begin{multline}\label{eq9.9}
\mathcal{Y} = (P_1-P_2)_{\mu} \sdot \langle S(P_{2}) | J^{\mu}| P(P_{1}) \rangle =   \\
f_{-}(q^{2}) q^2   +  
f_{+}(q^{2}) (M_{1}^2-M_{2}^2) \\ + B(q^{2})  P_1 \sdot \omega \; (1 - y)\ ,
\end{multline}
with the meson masses, $M_i$, and
\begin{multline}\label{eq9.10}
\mathcal{Z} = \frac{\omega_{\mu}  \sdot \langle S(P_{2}) | J^{\mu}| P(P_{1}) \rangle}{\omega \sdot P_{1}} = \\ 
f_{-}(q^{2}) ( 1 - y) + f_{+}(q^{2})
  ( 1 +  y)\ . 
\end{multline}
In Eqs.~(\ref{eq9.8},~\ref{eq9.9},~\ref{eq9.10}) the term $y$ defines the ratio between the two momenta $P_{1}$ and 
$P_{2}$ times the light-like four vector $\omega$ as, 
\begin{eqnarray}\label{eq9.11}
y = \frac{\omega \sdot P_{2}}{\omega \sdot P_{1}}= \frac{M_{2}^{2} + P_{1} \sdot P_{2}}{M_{1}^{2}+
 P_{1} \sdot P_{2}}\ , 
\end{eqnarray}
with $P_{1} \sdot P_{2} = \frac{1}{2}(M_{1}^{2}+M_{2}^{2}-q^{2}).$  For $q^{2} > 0$, 
it is convenient to restrict ourselves to the plane defined by 
 $\omega \cdot q = 0$. 
 This condition is allowed  in the system of reference where ${\bf P_{1}} + {\bf P_{2}} 
= \bf{0}$ with ${P_{1}}_{0} - {P_{2}}_{0} \neq 0$. 
 From the scalar
products $\mathcal{X}, \mathcal{Y}$ and $\mathcal{Z}$ we can isolate the form
factors $f_{\pm}(q^{2})$ from  $B(q^{2})$. Then, one gets the expressions for the  form factors:
\begin{equation}\label{eq9.12}
f_{\pm}(q^{2})= \Omega (y,q^{2}) \Psi_{\pm} (y,q^{2},\mathcal{X},\mathcal{Y},\mathcal{Z})\ ,
\end{equation}
where $\Omega (y,q^{2})$ is identical for both form factors $f_{\pm}(q^{2})$:
\begin{eqnarray}\label{eq9.13}
\vspace{-1.cm}
\Omega (y,q^{2}) =   \frac{1}{4 \bigl[ ((y-1) M_{1}^{2}+ q^{2}) y - M_{2}^{2} (y-1)\bigr]}\ ,
\end{eqnarray}
and where the functions $\Psi_{\pm} (y,q^{2},\mathcal{X},\mathcal{Y},\mathcal{Z})$ have the following forms:
\begin{multline}
\vspace{-0.8cm}
\Psi_{-} (y,q^{2},\mathcal{X},\mathcal{Y},\mathcal{Z}) =  \\
\mathcal{Y} (y+1)^{2}+\mathcal{X}  (y^{2}-1) + \\ \Bigr[(1-3 y) M_{1}^{2}- M_{2}^{2} (y-3)+  q^{2} (y-1)\Bigl]
 \mathcal{Z}\ , \nonumber 
\end{multline}
and
\begin{multline}
\vspace{-1.5cm}
\Psi_{+} (y,q^{2},\mathcal{X},\mathcal{Y},\mathcal{Z}) =  \\
 \mathcal{Y} (y^{2}-1) + \mathcal{X} (y-1)^{2}
+  \\ \Bigl[(y-1) M_{1}^{2} - M_{2}^{2} (y-1) + q^{2} (y+1)\Bigl] \mathcal{Z}\ .
\end{multline}
The second step is to express the amplitude $\langle S(P_{2}) | J^{\mu} | P(P_{1}) \rangle$ without using the 
form factors $f_{\pm}(q^{2})$. 
In CLFD the leading contribution to the transition amplitude $\langle S(P_{2}) | J^{\mu} | P(P_{1}) \rangle$ is 
given by the diagram shown in Fig.~2. 
\begin{figure}[b] 
\begin{center}
\includegraphics*[width=0.99\columnwidth]{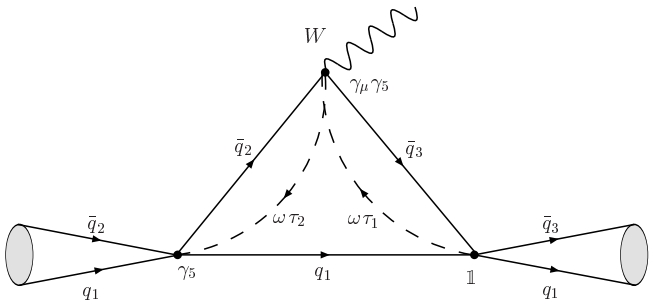}
\end{center}
\caption{Leading contribution in pseudoscalar-scalar transition.} 
\label{fig2} 
\end{figure}
By using the CLFD rules, one can derive the matrix elements from the diagram given in Fig.~2 
and  one has, 
\begin{multline}\label{eq9.15}
\langle S(P_{2}) | J^{\mu} | P(P_{1}) \rangle =  \\
 \int  \frac{{\rm d}^2{\bf R}_{\perp}{\rm d} x {\rm d}\theta}{2x(1-x)(2\pi)^3} \frac{1}{1-x^{\prime}}
{\rm Tr}  \biggl[ - {\Bar \vartheta_{s}} (m_1+ \SLASH k_{1}) \\
 \gamma^{\mu}\gamma^5  
(m_{2}+ \SLASH k_{2}) \vartheta_{p} (m_{3}- \SLASH k_{3})\Biggr]\ ,
\end{multline}
where $\vartheta_{j}$ is defined by:
\begin{eqnarray}
\vartheta_{p} \propto  \frac{1}{\sqrt{2}} A_p({\bf k}^{2}) \gamma^5\ ,  \vartheta_{s} \propto 
\frac{1}{\sqrt{2}} A_s({\bf k}^{2})\ . 
\end{eqnarray}
The indices $i=p,s$ of $\vartheta_{i}$ denote the wave functions
 referring  to the initial pseudoscalar and final scalar 
mesons,  respectively. Note  that  $x$ and $x^{\prime}$ are the fraction of the momentum carried by a 
quark $q_{3}$ (spectator quark) as given by: 
\begin{equation}\label{eq9.18}
x= \frac{\omega \sdot k_{3}}{\omega \sdot P_1}\ , \quad {\rm and}\quad x^{\prime} = \frac{\omega \sdot k_{3}}
{\omega \sdot P_2}\ .
\end{equation}
Now, one can replace the hadronic matrix element $\langle S(P_{2}) | J^{\mu} | P(P_{1}) \rangle$, which appears  
in the scalar products ${\mathcal X, Y, Z}$ defined in Eqs.~(\ref{eq9.8},~\ref{eq9.9},~\ref{eq9.10}), by the hadronic 
matrix elements $\langle S(P_{2}) | J^{\mu} | P(P_{1}) \rangle$ calculated by applying the CLFD diagrammatic 
rules and given in Eq.~(\ref{eq9.15}). 
Hence, by using Eq.~(\ref{eq9.12}) we are  able to compute the form factors $f_{\pm}(q^{2})$ as a function of $q^{2}$ and
this  over the whole  available  momentum range $0<q^{2}< q^{2}_{max}$.

\section{Numerical Application}
Introducing another set of form factors  $F_{0}(q^{2})$ and $F_{1}(q^{2})$, the relationship between 
the two sets of form factors is
\begin{align}\label{eq9.5}
F_{1}(q^{2})& = -f_{+}(q^{2})\ , \nonumber \\
F_{0}(q^{2})& = -f_{+}(q^{2}) - \frac{q^{2}}{M^{2}_{1}-M^{2}_{2}}f_{-}(q^{2})\ .
\end{align}
\begin{figure}[hp]
\begin{center}
\includegraphics*[width=0.95\columnwidth]{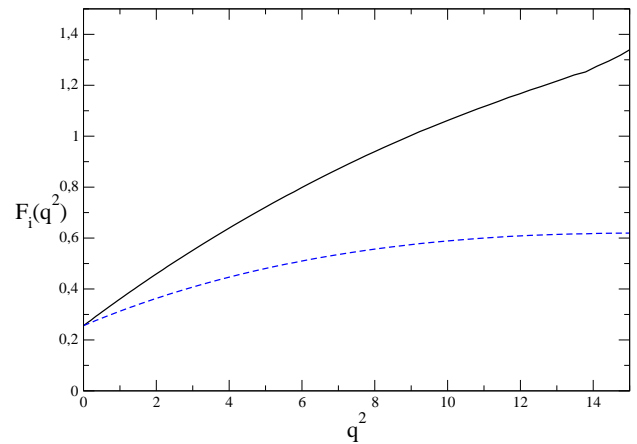}
\end{center}
\caption{Transition form factors  $F_0(q^2)$ (dashed line) and $F_1(q^2)$ (full line) 
 given  in case of $B^u \to f_0(980)^u$.} 
\label{fig3} 
\end{figure}
\begin{figure}[hp]
\begin{center}
\includegraphics*[width=0.95\columnwidth]{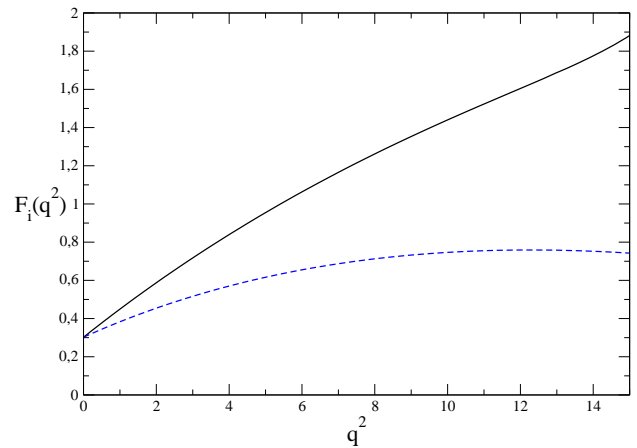}
\end{center}
\caption{Transition form factors  $F_0(q^2)$ (dashed line) and $F_1(q^2)$ (full line) 
given  in case of $B^s \to f_0(980)^s$.} 
\label{fig4} 
\end{figure} 

Note  that at $q^{2}=0$, one obtains $F_{1}(q^{2}=0)=F_{0}(q^{2}=0)=-f_{+}(q^{2}=0)$. We  
calculated the transition form factors in the case of the pseudoscalar scalar transitions, such as
 $B^u \rightarrow f_0(980)$ and $B^s \rightarrow f_0(980)$. The weak  transitions mentioned previously 
are induced by a current quark transition, $b \to u$ or $b \to s$ according to which component of 
the $f_0(980)$ one  focuses on. We are working in a constituent quark model 
where the CLFD formalism is applied. The wave functions used to describe the particles $B^u, B^s$ and $f_0(980)$ 
have been determined using the same approach as that for the form factors and  have been  constrained by 
physical observable such as decay constant and the normalization as well\cite{leitner:2006ab}. In our model 
of the scalar meson $f_0(980)$, one  assumes it is made of components $u\bar{u}, d\bar{d}$ and $s\bar{s}$ 
with constraints from $D$ 
branching ratios. 
All the details regarding its phenomenological determination can be found in paper\cite{leitner:2006ab}. 
The results are shown in Figs.~3 and~4 where the form factors $F_0(q^2)$ and $F_1(q^2)$ are plotted as a function 
of the momentum transfer $q^2$. Note that the theory allows us to directly obtain the behaviour of the form factors 
for all values of $q^{2}$ without extrapolation. Through the form factors $F_i(q^2)$ (that depend on the wave 
functions $B$ and $f_0(980)$), derived within the CLFD formalism, the hadronic matrix elements that drives 
the weak decay transition between two hadronic states can be known more precisely. The better we describe  
electroweak decay transitions, the better we will be able to understand quark flavour changing within the Standard Model.

\balance

\end{document}